 \documentclass[twocolumn,aps,prl,amsmath,amssymb,superscriptaddress]{revtex4-2}
\usepackage[colorlinks,linkcolor=blue,anchorcolor=blue,citecolor=blue,urlcolor=blue,filecolor=blue,menucolor=blue,runcolor=blue]{hyperref}
\usepackage{ulem}

\usepackage{graphicx}
\usepackage{multirow}
\usepackage{bm}
\usepackage{color}
\usepackage{ulem}
\usepackage{soul}
\usepackage{balance}
\usepackage{upgreek} 
\usepackage{subfigure}
\begin{document}


\title{Distinct pressure evolution of superconductivity and charge-density-wave in kagome superconductor CsV$_3$Sb$_5$ thin flakes}

\author{Ge Ye}
 \affiliation{Center for Correlated Matter, School of Physics, Zhejiang University, Hangzhou 310058, China}%
\author{Mengwei Xie}
 \affiliation{Center for Correlated Matter, School of Physics, Zhejiang University, Hangzhou 310058, China}%
\author{Chufan Chen}
\affiliation{Center for Correlated Matter, School of Physics, Zhejiang University, Hangzhou 310058, China}%
\author{Yanan Zhang}
\affiliation{Center for Correlated Matter, School of Physics, Zhejiang University, Hangzhou 310058, China}%
\author{Dongting Zhang}
\affiliation{Center for Correlated Matter, School of Physics, Zhejiang University, Hangzhou 310058, China}%
\author{Xin Ma}
\affiliation{Center for Correlated Matter, School of Physics, Zhejiang University, Hangzhou 310058, China}%
\author{Xiangyu Zeng}
\affiliation{Hangzhou Institute of Technology, Xidian University, Hangzhou, 311200, China}%
\author{Fanghang Yu}
\affiliation{CAS Key Laboratory of Strongly-coupled Quantum Matter Physics, Department of Physics, University of Science and Technology of China, Hefei, 230022, China}
\author{Yi Liu}
\affiliation{Key Laboratory of Quantum Precision Measurement of Zhejiang Province, Department of Applied Physics, Zhejiang University of Technology, Hangzhou, 310014, China}
\author{Xiaozhi Wang}
\affiliation{College of Information Science and Electronic Engineering, Zhejiang University, Hangzhou, 310058, China}
\author{Guanghan Cao}
\affiliation{School of Physics, Zhejiang University, Hangzhou, 310058, China}
\affiliation{State Key Laboratory of Silicon and Advanced Semiconductor Materials, Zhejiang University, Hangzhou, 310058, China}
\author{Xiaofeng Xu}
\affiliation{Key Laboratory of Quantum Precision Measurement of Zhejiang Province, Department of Applied Physics, Zhejiang University of Technology, Hangzhou, 310014, China}
\author{Xianhui Chen}
\affiliation{CAS Key Laboratory of Strongly-coupled Quantum Matter Physics, Department of Physics, University of Science and Technology of China, Hefei, 230022, China}
\author{Huiqiu Yuan}
\affiliation{Center for Correlated Matter, School of Physics, Zhejiang University, Hangzhou 310058, China}
\affiliation{State Key Laboratory of Silicon and Advanced Semiconductor Materials, Zhejiang University, Hangzhou, 310058, China}
\author{Chao Cao}
\email{ccao@zju.edu.cn}
\affiliation{Center for Correlated Matter, School of Physics, Zhejiang University, Hangzhou 310058, China}
\author{Xin Lu}%
 \email{xinluphy@zju.edu.cn}
\affiliation{Center for Correlated Matter, School of Physics, Zhejiang University, Hangzhou 310058, China}%

\date{\today}

\begin{abstract}
  It is intriguing to explore the coexistence and (or) competition between charge-density-wave (CDW) and superconductivity (SC) in many correlated electron systems, such as cuprates\cite{PhysRevB.83.104506,PhysRevLett.99.067001,PhysRevB.79.100502}, organic superconductors\cite{PhysRevB.100.060505,PhysRevB.80.092508} and dichacolgenides\cite{nnano.2015.143,PhysRevB.87.094502}. Among them, the recently discovered $\mathbb{Z} _2$ topological kagome metals $A$V$_3$Sb$_5$ ($A$=K, Rb, Cs) serve as an ideal platform to study the intricate relation between them. Here, we report the electrical resistance measurements on CsV$_3$Sb$_5$ thin flakes ($\approx$ 60 nm) under hydrostatic pressure up to 2.12 GPa to compare its pressure phase diagram of CDW and SC with its bulk form. Even though the CDW transition temperature ($T_{\mathrm{CDW}}$) in CsV$_3$Sb$_5$ thin flakes is still monotonically suppressed under pressure and totally vanishes at $P_2$=1.83 GPa similar to the bulk, the superconducting transition temperature ($T_c$) shows an initial decrease and consequent increase up to its maximum $\sim$ 8.03 K at $P_2$, in sharp contrast with the M-shaped double domes in the bulk CsV$_3$Sb$_5$. Our results suggest the important role of reduced dimensionality on the CDW state and its interplay with the SC, offering a new perspective to explore the exotic nature of CsV$_3$Sb$_5$. 
  
\end{abstract}

\maketitle

The recently discovered kagome family $A$V$_3$Sb$_5$ ($A$=K, Rb, Cs) with a frustrated geometry have attracted great interest due to the rich interplay between electronic correlations, topology, unconventional charge-density-wave (CDW), and superconductivity (SC) \cite{PhysRevMaterials.3.094407,PhysRevMaterials.5.034801,PhysRevLett.125.247002,Yin_2021}. All members in the $A$V$_3$Sb$_5$ family show a superconducting state below $T_c$ of 2.5, 0.92, and 0.93 K for $A$=Cs, Rb, and K, respectively \cite{PhysRevLett.125.247002,Yin_2021,PhysRevMaterials.5.034801}, while the corresponding charge order emerges at $T_{\mathrm{CDW}}$ of 94, 102, 78 K, respectively \cite{PhysRevLett.126.247001,PhysRevX.11.031050,PhysRevB.103.L220504}. Scanning tunneling microscopy (STM)\cite{STM_KV3Sb5, nature_STM_Cs, Nature_STM_GaoHongjun, PhysRevX.11.031026}, X-ray diffraction (XRD) \cite{PhysRevX.11.031050} and ARPES \cite{ARPES_CS_CDWgap_Gaohongjun, ARPES_Cs_CDWgap_Taka, ARPES_Rb_CDWgap_Wangshancai} experiments have all revealed a $2\times2$ charge modulation in the V-based kagome lattice layers below $T_{\mathrm{CDW}}$. Further ARPES \cite{PhysRevB.106.L241106}, XRD \cite{PhysRevX.11.041030,PhysRevB.105.195136} and NMR measurements \cite{s41535-022-00437-7} have proposed a $2\times2\times2$ or $2\times2\times4$ CDW order with the stacking of star-of-David (SoD) or tri-hexagonal (TrH) distortions along the \textit{c} axis, favoring a 3D nature of the CDW due to interlayer couplings. Additional breaking of time-reversal and rotational symmetry were also reported in the CDW state and an electronic origin of the CDW state has been claimed to be probably associated with the Fermi surface nesting between van Hove saddle-points at M points \cite{s41586-021-04327-z,PhysRevResearch.4.023244,PhysRevMaterials.5.L111801}. 

\begin{figure}[htb]
  \centering
  \includegraphics[angle=0,width=0.45 \textwidth]{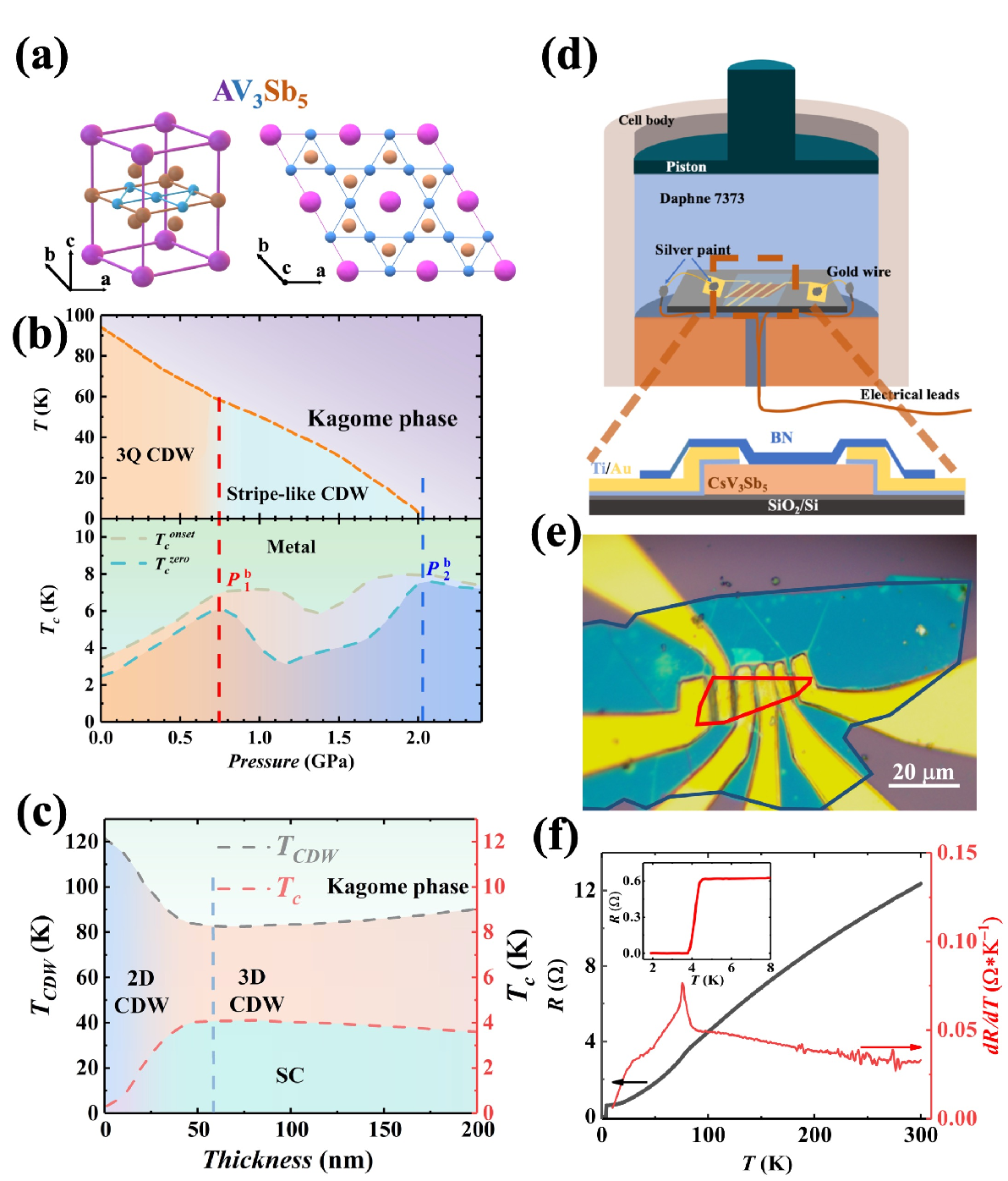}
	\vspace{0pt}
  \caption{\textbf{Schematic illustration of crystal structure and phase diagrams of CsV$_3$Sb$_5$ and the experimental set-ups for electrical resistance measurements on CsV$_3$Sb$_5$ thin flake under pressure.} \textbf{a,} Three-dimensional crystal structure of CsV$_3$Sb$_5$ on the left with its top view on the right. \textbf{b,} Phase diagrams for the bulk CsV$_3$Sb$_5$ under pressure with the CDW $T_{\mathrm{CDW}}$ (top), SC $T_c^{\mathrm{onset}}$ and $T_c^{\mathrm{zero}}$ (bottom) adapted from Refs. \cite{PhysRevLett.126.247001,NC_Phasediagram,s41586-022-05351-3}. \textbf{c,} Thickness dependence of $T_{\mathrm{CDW}}$ and $T_c$ for CsV$_3$Sb$_5$ flakes with the blue dashed line as the crossover region, adapted from Refs. \cite{Lishiyan}. \textbf{d,} Schematic illustration of the experimental set-ups, where the CsV$_3$Sb$_5$ thin flake device was encapsulated into a piston cell with Daphne 7373 as the pressure medium. The bottom of \textbf{d} shows a section view of the device structure. \textbf{e,} Optical microscopy image of the device. The CsV$_3$Sb$_5$ thin flake and $h$-BN are marked with red and blue solid lines, respectively. 
  \textbf{f,} Temperature dependence of electrical resistance (black) R(T) and its derivative (red) $dR/dT$ of the CsV$_3$Sb$_5$ thin flake device. The inset shows the SC transition at low temperatures in the R(T) curve. 
  }
  \label{fig1}
\end{figure}

In particular, the Cs compound has the highest $T_c$ at ambient pressure and displays a complex superconducting phase diagram compared with other members, where the bulk CsV$_3$Sb$_5$ shows a monotonic decrease of $T_{\mathrm{CDW}}$ till 2 GPa, and a puzzling M-shaped double-dome feature of superconductivity where the $T_c$ peaks at $P{^{b}_1}$ $\sim$ 0.7 GPa and $P{^{b}_2}$ $\sim$ 2.0 GPa, respectively, as shown in Fig.\ref{fig1}(b) \cite{PhysRevLett.126.247001,NC_Phasediagram}. Recent NMR experiments \cite{s41586-022-05351-3} argued that a stripe-like CDW order should replace the original one at $P{^{b}_1}$ with domain walls and be totally suppressed at $P{^{b}_2}$, yielding a double-dome SC in the pressure phase diagram. On the other hand, the CsV$_3$Sb$_5$ crystals can be exfoliated to thin flakes of a few nanometers thickness, and its CDW and superconducting state show an opposite but non-monotonic evolution with flake thickness \cite{PhysRevLett.127.237001,Lishiyan}: $T_c$ is enhanced to the maximum value of 4.5 K around 30 to 60 nm and decreases to 0.9 K at 5 nm, while its non-monotonic $T_{\mathrm{CDW}}$ of the CDW order signals a possible 3D to 2D crossover around 60 nm. Recently, hydrostatic pressure has been successfully applied in a variety of two-dimensional materials to tune their quantum ground states, such as twisted bilayer graphene\cite{doi:10.1126/science.aav1910}, transition metal dichalcogenides \cite{MoS2_pressure_2015,MoS2_NanoLett}, graphene/$h$-BN Moir$\acute{e}$ superlattices \cite{BN_Graphene_moire_Nature}, and CrI$_3$ \cite{Pressure_CrI3}. In this article, we report our electrical transport measurements of mechanically exfoliated CsV$_3$Sb$_5$ thin flakes under hydrostatic pressure, to systematically compare the behaviors between the bulk and thin flakes and unveil a common mechanism of tuning the CDW and SC in CsV$_3$Sb$_5$ by pressure and dimensionality.

Several CsV$_3$Sb$_5$ thin flake devices ($\approx$ 60 nm) on SiO$_2$/Si substrate have been fabricated with Ti/Au electrodes for electrical resistance measurements and encapsulated in a piston pressure cell. CsV$_3$Sb$_5$ thin flakes were obtained by mechanical exfoliation and all devices were prepared by the standard photolithography process. The flake thickness could be roughly estimated from its optical color in microscope and precisely confirmed by scanning AFM in tapping-mode. CsV$_3$Sb$_5$ flakes were transferred onto the Si substrate (290 nm SiO$_2$) by the polydimethylsiloxane (PDMS) transfer method. Ti/Au (5 nm/50 nm) films were deposited by magnetron sputtering in sequence as the electrodes for better electrical contacts. Finally, the exfoliated $h$-BN flakes were transferred onto the devices by the PDMS transfer method as a capping layer to protect the device from further degradation during operations. The schematic figure of the experimental set-up is shown in Fig. \ref{fig1}(d) and the CsV$_3$Sb$_5$ thin flake and h-BN capping layer in the device are marked with red and blue lines, respectively, in Fig.\ref{fig1}(e). All the transfer processes were performed in the glove box with both oxygen and water vapor content lower than 10 ppm to avoid oxidization. 

The device was diced into 1.5 mm $\times$ 1.5 mm in size to fit the inner bore of the pressure cell, as shown in Fig.\ref{fig1}(d). The whole device was encapsulated in a piston-cylinder-type cell with Daphne 7373 as the pressure transmitting medium and the pressure at low temperature was calibrated by the Pb superconducting transition temperature from electrical resistivity measurements. PPMS and Oxford $^3$He refrigerators were used to cool devices down to 1.8 K and 0.3 K, respectively, with magnetic field up to 14 T along the \textit{c} direction. Since the device is fragile to static charge or current pulse during preservation or measurement, special cares with proper grounding should be called for. 

In order to cross-check the results, high quality and well-characterized CsV$_3$Sb$_5$ single crystals were obtained from two different groups. Several CsV$_3$Sb$_5$ thin flakes from each group were prepared and they indeed show a consistent behavior under pressure. For one device, the thin CsV$_3$Sb$_5$ flake is calibrated 58.6 nm in thickness, where its $T_{\mathrm{CDW}}$ and $T_c$ reaches the minimum and maximum value, respectively \cite{Lishiyan}. Fig.\ref{fig1}(f) shows the temperature-dependent resistance and its derivative ($dR/dT$) for the CsV$_3$Sb$_5$ thin flake at ambient pressure, where the kink anomaly in resistance and the corresponding peak in $dR/dT$ signals a CDW transition at $T_{\mathrm{CDW}}$ = 76 K and a sharp superconducting transition can be observed with $T_c^{\mathrm{onset}}$ = 4.4 K and $T_c^{\mathrm{zero}}$ = 3.8 K as in the inset of Fig.\ref{fig1}(f). Our $T_{\mathrm{CDW}}$ and $T_c$ values are consistent with the previous report for 60 nm CsV$_3$Sb$_5$ thin flakes \cite{Lishiyan}. 

\begin{figure*}[http]
\includegraphics[angle=0,width=1.05\textwidth]{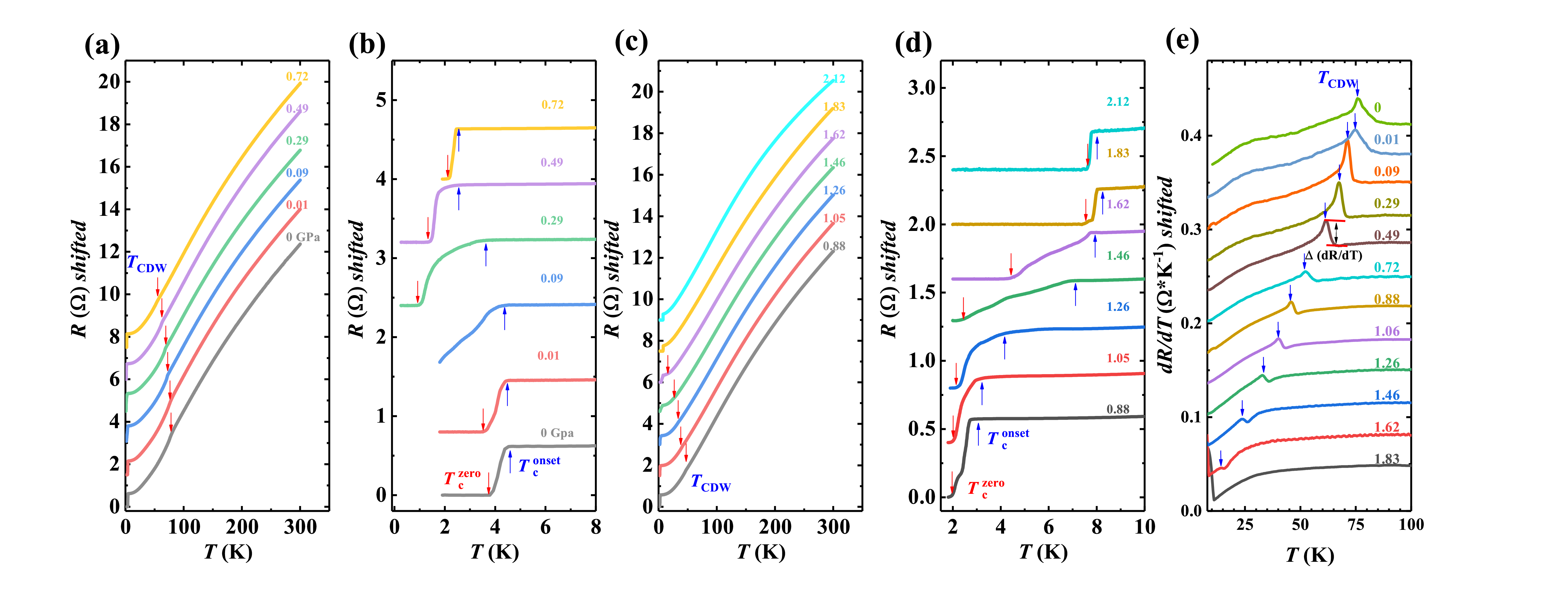}
\vspace{-25pt}
\caption{\textbf{Temperature dependence of the electrical resistance for the CsV$_3$Sb$_5$ thin flake device under a series of pressures. }
\textbf{a, c,} Temperature-dependent resistance for device No. 7 in the low and high pressure range, respectively. 
The CDW transition temperatures $T_{\mathrm{CDW}}$ are marked by red arrows. \textbf{b, d,} SC transition temperatures for the same device in the respective low and high pressure ranges, with blue and red arrows mark the $T_c^{\mathrm{onset}}$ and $T_c^{\mathrm{zero}}$, respectively. \textbf{e,} The derivative of resistance curves, $dR/dT$, under various pressures. The peaks in $dR/dT$ are marked by blue arrows, signaling the CDW $T_{\mathrm{CDW}}$. All curves are vertically shifted for clarity. 
}
  \label{fig2}

\end{figure*}

\begin{figure}[http]
  \includegraphics[angle=0,width=0.48\textwidth]{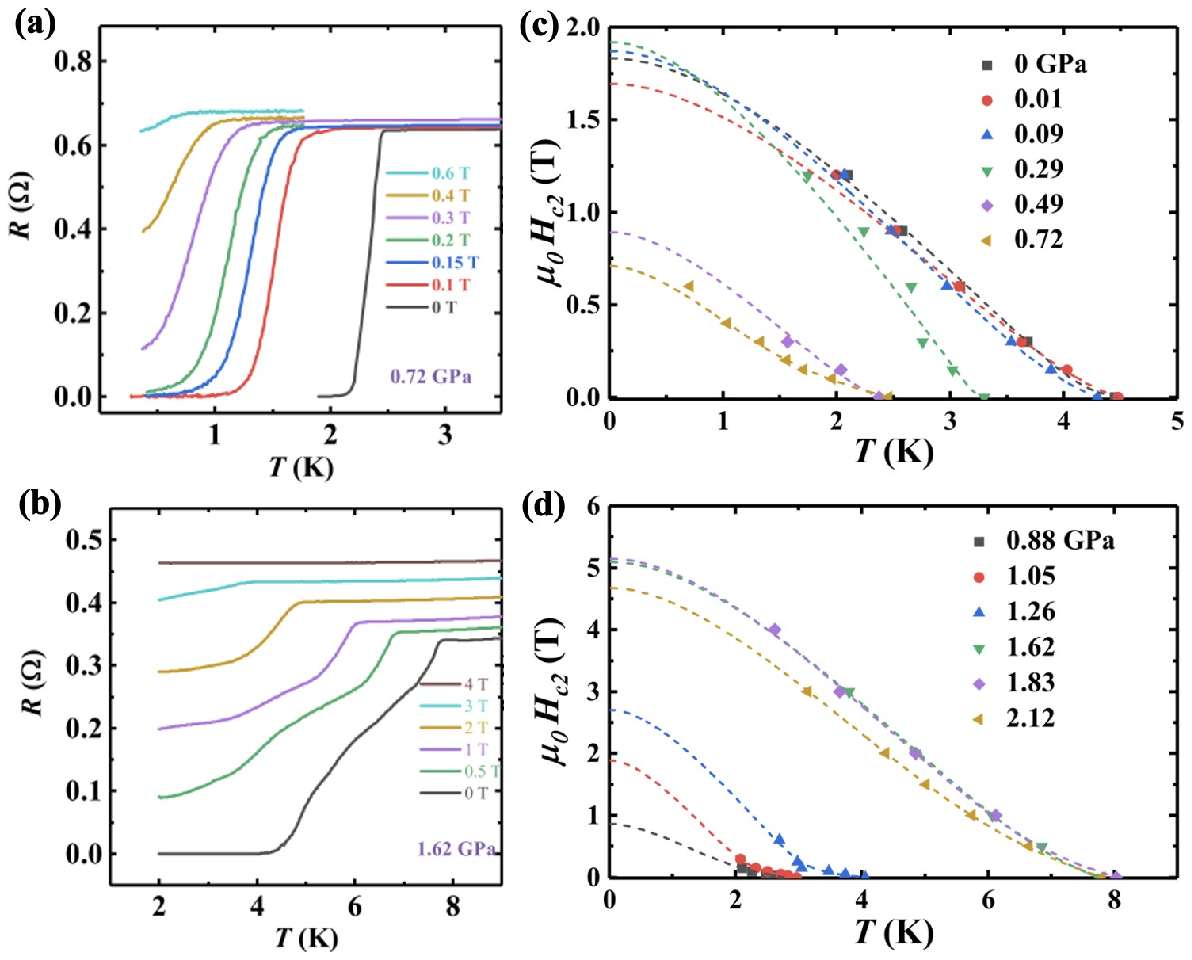}
  \vspace{-10pt}
  \caption{\textbf{Upper critical field for device No. 7 under different pressures.}
  \textbf{a, b}, Temperature dependence of resistance under magnetic field applied along c-axis at $P$ = 0.72, 1.62 GPa, respectively. \textbf{c, d}, The evolution of $H_{c2}$-$T$ phase diagram (determined by $T_{c}^{onset}$) as a function of pressure. The dashed lines are two band model fitting curves\cite{PhysRevB.67.184515}.
    }
    \label{fig3}
  \end{figure}

Figure \ref{fig2}(a) and (c) show the temperature-dependent electrical resistance for the CsV$_3$Sb$_5$ thin flake under a series of pressures up to 2.12 GPa, illustrating the systematical evolution of SC and CDW states as a function of pressure. Similar to the bulk case, the $T_{\mathrm{CDW}}$ in CsV$_3$Sb$_5$ can be determined by the kink structure in the resistance curves $R(T)$ or the peak-like anomaly in the derivative resistance curves $dR/dT$, as in Fig.\ref{fig2}(e). With increased pressure, the $T_{\mathrm{CDW}}$ is smoothly suppressed and reaches zero at a pressure of $P_2$ = 1.83 GPa, comparable to but slightly smaller than $P_2^b\sim$2.0 GPa in the bulk. Moreover, we notice a sudden change of the peak feature in $dR/dT$ from a broad peak at ambient pressure to a sharp peak-dip structure at 0.09 GPa as in Fig.\ref{fig2}(e), and that the peak gets suppressed but the dip becomes more prominent with increased pressure until 1.83 GPa. We ascribe the change in $dR/dT$ at 0.09 GPa in the thin flake to the same origin as those for features observed at $P^b_1\sim$ 0.7 GPa in the bulk, signaling a possible transition of the CDW order. It seems that the reduced thickness of the CsV$_3$Sb$_5$ thin flake has tuned the emergent CDW order at $P^b_1$ to be more energetically favored in a much lower pressure at $P_1\sim$ 0.09 GPa.

\begin{figure}[ht]
  \centering
  \includegraphics[angle=0,width=0.49 \textwidth]{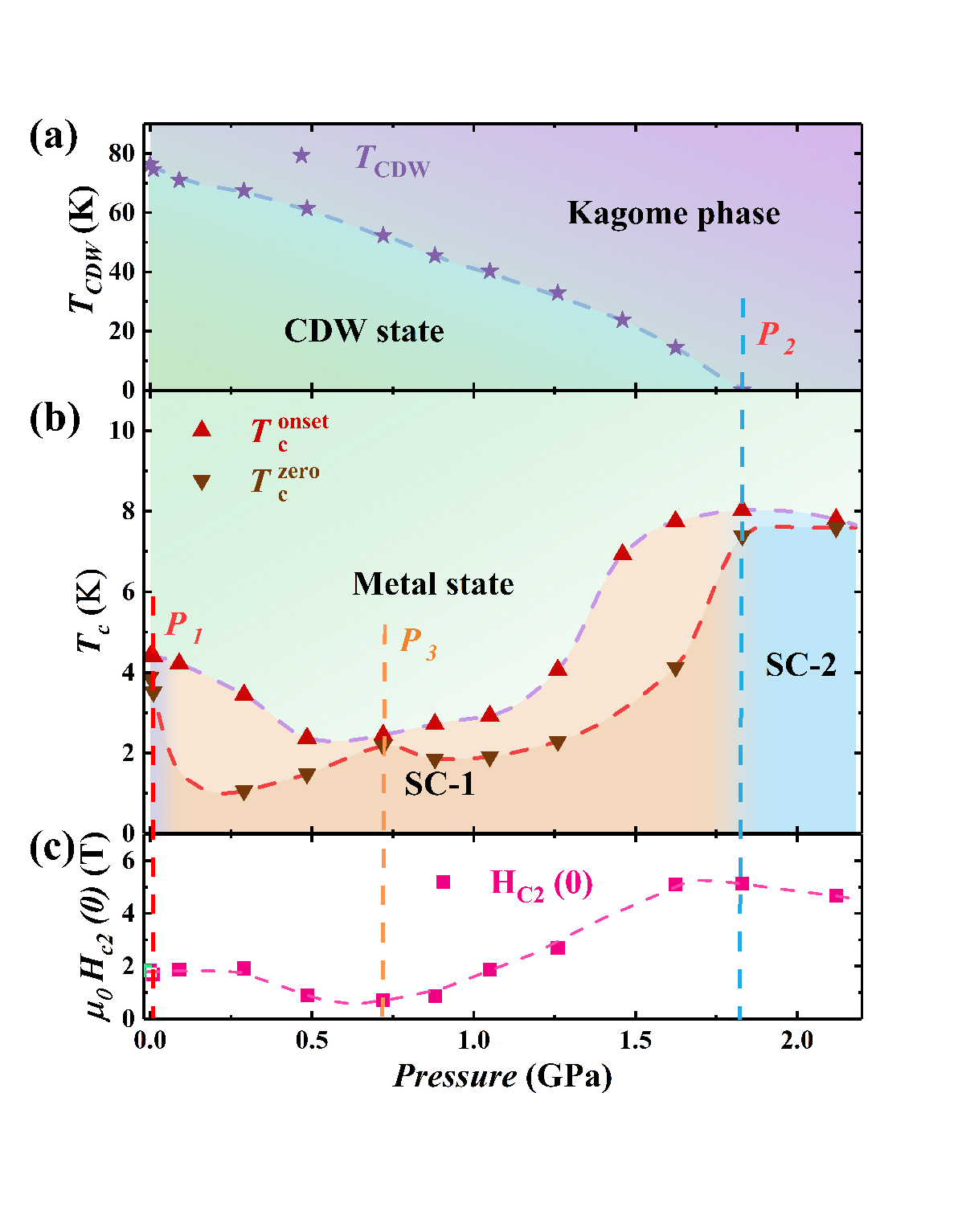}
	\vspace{-50pt}
 \caption{\textbf{The pressure-temperature phase diagram of CsV$_3$Sb$_5$ thin flakes.} 
 \textbf{a,} Pressure dependence of CDW transition temperatures. CDW is monotonically suppressed and vanishes at $P_2$. 
 \textbf{b,} Pressure dependence of SC transition temperatures $T_c^{\mathrm{zero}}$ and $T_c^{\mathrm{onset}}$ from R(T) curves. 
 \textbf{c,} The SC upper critical field $\mu_{0}H_{c2}(0)$ as a function of pressure. $\mu_{0}H_{c2}(0)$ is obtained by the two-band model fitting. 
 }
 \label{fig4}
\end{figure} 

Figure \ref{fig2}(b) and (d) show the enlarged view with focus on the SC evolution in the CsV$_3$Sb$_5$ thin flake under pressure, where the red and blue arrows mark the onset and zero-resistance transition temperatures, $T_c^{\mathrm{onset}}$ and $T_c^{\mathrm{zero}}$, respectively. If we track the evolution of $T_c^{\mathrm{onset}}$ for the CsV$_3$Sb$_5$ thin flake as a function of pressure, it shows an immediate decrease upon squeezing, opposite to an initial increase of $T_c^{\mathrm{onset}}$ to a local maximum around 0.7 GPa in the bulk, and becomes roughly constant around 2.5 K between 0.49 and 1.05 GPa before ramping up to another local maximum of 8 K at 1.83 GPa, similar to the observation in the bulk CsV$_3$Sb$_5$, as shown in Fig.\ref{fig2}(b). When comparing the superconducting behaviors between the thin flake and bulk, two local maximum $T_c$ pressures for the double superconducting domes in the bulk CsV$_3$Sb$_5$ at $P^b_1\sim$0.7 GPa and $P^b_2\sim$ 2.0 GPa have shifted to $P_1 <$0.10 GPa and $P_2\sim$1.8 GPa, respectively. In particular, the shifted $P_1$ value to less than 0.1 GPa also corroborates our claim that the original CDW order switch has now been tuned to 0.09 GPa in the thin flake, assuming a stronger competition between the emergent CDW and SC between $P_1$ and $P_2$. Above $P_2$, $T_c^{\mathrm{onset}}$ decreases together with $T_c^{\mathrm{zero}}$ as the CDW state vanishes. 

We stress that the superconducting transitions are very broad between $P_1$ and $P_2$, exhibiting either multiple-step or long-tail features. For example, $T_c^{\mathrm{onset}}$ and $T_c^{\mathrm{zero}}$ are 3.5 and 1.0 K, respectively, at 0.29 GPa, while they are 7.8 and 4.1 K at 1.62 GPa. The transition-width is comparable to $T_c^{\mathrm{onset}}$ and such a broad transition generally implies an inhomogeneous superconductivity in the flake. Since the transition becomes sharp again once the CDW order is suppressed at 1.83 GPa, extrinsic origins such as pressure inhomogeneity can be excluded. We argue that the SC inhomogeneity arises from the existence of different CDW domains with enhanced scatterings as in the case of bulk CsV$_3$Sb$_5$. However, the SC transition at $P_3$ = 0.72 GPa in ($P_1$, $P_2$) is sharp with $\Delta T_c=T_c^{\mathrm{onset}}-T_c^{\mathrm{zero}}=$0.3 K, and such behavior is absent in the bulk CsV$_3$Sb$_5$, suggesting a homogeneous CDW order at 0.72 GPa for the thin flake. 

Figure \ref{fig3} illustrates the suppression of superconductivity under an applied magnetic field along the c-axis for CsV$_3$Sb$_5$ thin flakes, where the superconducting transition gets widened under magnetic field, indicative of increased inhomogeneity of SC. To study the zero-temperature upper critical field ($\mu_0 H_{c2}(0)$), and its evolution with pressure, the onset of superconductivity in the field was defined as its $T_c$ and its dependence with temperature was fit using a two-band model\cite{PhysRevB.67.184515}. An enhancement in both $\mu_0 H_{c2}(0)$ and $T_c$ around 1.8 GPa corresponds to the state in the absence of CDW order. Additionally, a sharp superconducting transition and a local minimum of $\mu_0 H_{c2}(0)$ at 0.72 GPa suggest a competitive yet coexisting relationship between the superconducting state and CDW order.

The pressure-dependent phase diagrams of CDW ($T_{\mathrm{CDW}}$) and SC ($T_c^{\mathrm{onset}}$ and $T_c^{\mathrm{zero}}$) in CsV$_3$Sb$_5$ thin flakes are plotted in Fig.\ref{fig4}(a) and (b), based on the electrical resistance measurements under pressure, in comparison with its bulk form as in Fig.\ref{fig1}(b). Even though the CDW order still decreases monotonically with pressure and completely disappears at $P_2 \sim$ 1.8 GPa with a similar $T_c$ maximum as in the bulk, the SC in CsV$_3$Sb$_5$ thin flakes show a remarkably different behavior. The bulk CsV$_3$Sb$_5$ displays an M-shaped double superconducting domes and a shallow $T_c$ valley in between, with its two local $T_c$ maximum at $P^b_1\sim$ 0.7 GPa and $P^b_2\sim$ 2.0 GPa, respectively. For CsV$_3$Sb$_5$ thin flakes, however, the higher-pressure SC dome is barely changed while the lower-pressure SC dome centered around 0.7 GPa in the bulk has been lowered to the ambient pressure, $P_1 <$ 0.1 GPa. As in Fig.\ref{fig4}(b), $T_c^{\mathrm{zero}}$ decreases rapidly above ambient pressure while $T_c^{\mathrm{onset}}$ changes little in a small pressure range of 0.1 GPa, which resembles well with the bulk around $P^b_1\sim$ 0.7 GPa. We argue that the original CDW switch at 0.7 GPa has shifted to below 0.1 GPa due to the reduced thickness in the thin flake, and causes a change of the SC phase diagram in accordance \cite{PhysRevLett.125.247002,NC_Phasediagram}. Moreover, the superconducting $T_c^{\mathrm{onset}}$ and $T_c^{\mathrm{zero}}$ are connected at $P_1$ and $P_2$, and form a loop, which is divided into two separate lobes by a node at $P_3 \sim$ 0.72 GPa. In each lobe, the presence of CDW domains yield an inhomogeneous SC with a broad transition due to the strong competition between SC and CDW as in the case of 1$T$-TaSe$_2$ \cite{10.1038/nphys2935}. When magnetic field is applied perpendicular to the CsV$_3$Sb$_5$ thin flake, its upper critical field $\mu_{0}H_{c2}(0)$ as a function of pressure is plotted in Fig.\ref{fig4}(c), showing a weakened double-dome structure compared with the bulk (The temperature-dependent resistance curves are included in Fig. \ref{fig3} (c, d) for different fields and the $\mu_{0}H_{c2}(0)$ at zero Kelvin can be obtained by a two-band model fitting\cite{PhysRevB.67.184515}).

\begin{figure}[ht]
  \centering
  \includegraphics[angle=0,width=0.45 \textwidth]{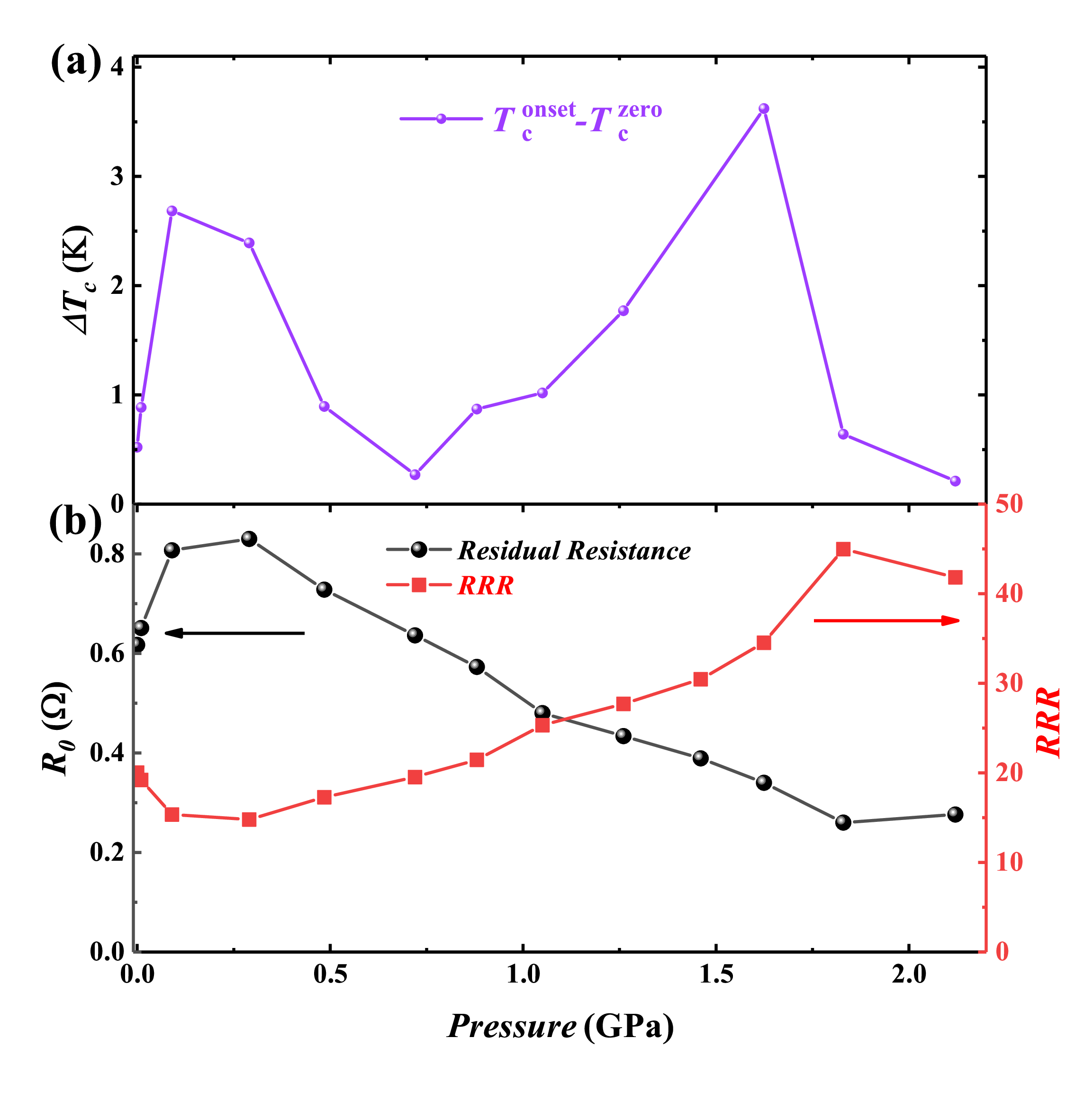}
	\vspace{-20pt}
  \caption{\textbf{Pressure dependence of $\Delta (T_c)$, residual resistance and residual-resistance ratio for the CsV$_3$Sb$_5$ thin flake.} 
  \textbf{a,} $\Delta (T_c)$ as a function of pressure (determined by $T_c^{\mathrm{onset}}$ - $T_c^{\mathrm{zero}}$ for device No. 7). 
  \textbf{b,} Pressure dependence of residual resistance and residual-resistance ratio (RRR), where the residual resistance is defined by the resistance value at 10 K for the specific pressure. 
  }
  \label{fig5}
\end{figure}

Our experimental results can be explained by the tuning of CDW orders with reduced dimensionality in the CsV$_3$Sb$_5$ thin flake. The CDW in CsV$_3$Sb$_5$ at ambient pressure has been reported to evolve from 3D to 2D in nature with reduced sample thickness as in Fig.\ref{fig1}(c), where the vanishing of the $A_{\mathrm{1g}}$ Raman frequency shift is observed below a critical thickness of 25 layers, signaling a weakened electron-phonon coupling\cite{Lishiyan}. For the CsV$_3$Sb$_5$ thin flake, reduced thickness would reduce the interlayer coupling and suppress the 3D CDW ordering along the c axis, but friendly to the 2D CDW instead. On the other hand, for the bulk, recent NMR results claim that the 3$Q$-CDW state should be dominant at the pressure below $P^b_1$, with a 3D character due to interlayer couplings, and transforms into a 2D stripe-like CDW with a unidirectional 4$a_0$ modulation under a moderate pressure in ($P^b_1$, $P^b_2$), as in Fig.\ref{fig1}(b) \cite{s41586-022-05351-3}. The 3$Q-$ and stripe-like CDWs should be very close in energy and compete for the electronic ground state in CsV$_3$Sb$_5$\cite{s41586-022-05351-3}, and the vertical Cs-Sb atomic distance is more sensitive to the hydrostatic pressure, resulting in a dominant reduction of the $c/a$ ratio\cite{PhysRevLett.126.247001}. In such a case, the 3$Q$ CDW may lift its energy due to interlayer couplings and yield to the stripe-like CDW around $P^b_1$ \cite{10.21468/SciPostPhys.12.2.049,Hasan_review}. We speculate that the 2D CDW in CsV$_3$Sb$_5$ thin flake and stripe-like CDW for the bulk should be identical in origin and antagonistic to the SC, causing the SC valley in pressure phase diagrams. In such a scenario, the dimensionality and hydrostatic pressure may play the same role in tuning the ground state of CsV$_3$Sb$_5$. Even though our first-principle calculation on the phonon modes for the bulk CsV$_3$Sb$_5$ under pressure and a bilayer slab can not explicitly identify the CDW transition, it suggests the delicate competition between different CDW orders with different stacking modulation and a robust feature of the in-plane distortion(see section 4 in Supplemental Material)\cite{arxiv_Son}. Because our CsV$_3$Sb$_5$ thin flake is close to crossover  thickness from 3D to 2D CDW, it is natural to expect the CDW switch would be lowered to only $<0.1$ GPa under pressure and the corresponding SC $T_c$ peak should be observed at the new $P_1$. 

We note that STM measurements have reported a stripe-like charge order with $4a_0/3$ spatial modulation on the surface of bulk CsV$_3$Sb$_5$ and its relation with the stripe-like CDW between ($P^b_1$, $P^b_2$) is not clear yet \cite{nature_STM_Cs,Nature_STM_GaoHongjun}. Moreover, the $T_c$ of CsV$_3$Sb$_5$ thin flakes below 1.8 GPa is relatively lower than the bulk, indicating a more intense competition between SC and 2D CDW with reduced sample thickness. The coexistence of stripe-like CDW and SC has also been observed in the high-$T_c$ cuprate superconductors La$_{2-x}$Ba$_x$CuO$_4$, and pair-density-wave (PDW) SC has been proposed with non-zero momentum for Cooper pairs \cite{PhysRevLett.99.067001,PhysRevB.83.104506}. For other layered superconductors such as Ising superconductor NbSe$_2$, its $T_c$ decreases from 7.2 K in the bulk to 3 K for monolayer with reduced sample thickness, while its $T_{\mathrm{CDW}}$ shows an opposite increase from 33 K to 145 K\cite{nnano.2015.143,nnano.2012.193,nchem.1589,nphys3538}. However, when CDW is suppressed in NbSe$_2$ by increasing the interlayer distance, the $T_c$ seems robust against the change of sample thickness\cite{s41567-022-01778-7}. We observe that the $T_c$ for CsV$_3$Sb$_5$ thin flakes at $P_2$ (where CDW vanishes) is also comparable to the bulk value. We thus argue the variation of $T_c$ with thickness is probably due to the thickness-modulated CDW competing with superconductivity. It is intriguing to explore possible PDW SC at ambient pressure and the pressure phase diagram for thinner CsV$_3$Sb$_5$ flakes to check whether the SC $T_c$ can be recovered back to the bulk value when CDW is totally suppressed\cite{Lishiyan}. 

In summary, we have systematically investigate electrical resistance of the CsV$_3$Sb$_5$ thin flakes (58nm) under pressure to compare its pressure phase diagram with the bulk. In contrast to the M-shaped double SC domes, the CDW switch and first SC $T_c$ dome have shifted to below 0.1 GPa in the thin flake from the initial $P^b_1\sim$ 0.7 GPa as in the bulk. We would thus conclude that the hydrostatic pressure or sample thickness can both tune the subtle competition and balance between the 3D and 2D CDWs in CsV$_3$Sb$_5$ due to the interlayer interactions and the SC response to the CDW switch accordingly. Considering the complex relationship between CDW and SC competition in CsV$_3$Sb$_5$ thin flakes, we believe it is necessary to conduct low-temperature Raman spectroscopy under pressure for further studies, as in the case of 2H-TaS$_2$\cite{PhysRevLett.122.127001}. Furthermore,in order to shed more light on the nature of 3D and 2D CDW, it is worth to explore the evolution of CDW and SC for monolayer or multilayer CsV$_3$Sb$_5$ under pressure in the future.

\begin{acknowledgments}
This work was supported by the National Key R\&D Program of China(Grant No. 2022YFA1402200), the National Natural Science Foundation of China (Grant No. 12174333 and 12274364), and the Key R\&D Program of Zhejiang Province, China (Grant No. 2021C01002). X.X. acknowledges the financial support from National Natural Science Foundation of China (Grants No.12274369 and No. 11974061).
\end{acknowledgments}

\end{document}